\documentclass{webofc}
\usepackage[varg]{txfonts}   
\usepackage{enumitem}
\usepackage{booktabs}
\usepackage{setspace}
\newcommand{\eg}{\textit{e.g.}}
\newcommand{\ie}{\textit{i.e.}}

\newcommand{\yz}{{Y|Z}}
\begin{document}
\title{Forecasting the $Y_{500}-M_{500}$ scaling relation from the NIKA2 SZ Large Program}
%
%

\author{
    \firstname{F.} \lastname{K\'eruzor\'e} \inst{\ref{LPSC}}
    \fnsep\thanks{\email{keruzore@lpsc.in2p3.fr}}
    \and
    \firstname{E.} \lastname{Artis} \inst{\ref{LPSC}}
    \and
    \firstname{J.-F.} \lastname{Mac\'ias-P\'erez} \inst{\ref{LPSC}}
    \and
    \firstname{F.} \lastname{Mayet} \inst{\ref{LPSC}}
    \and
    \firstname{M.} \lastname{Mu\~noz-Echeverr\'ia} \inst{\ref{LPSC}}
    \and
    \firstname{L.} \lastname{Perotto} \inst{\ref{LPSC}}
    \and
    \firstname{F.} \lastname{Ruppin} \inst{\ref{MIT}}
}

\institute{
    Univ. Grenoble Alpes, CNRS, Grenoble INP, LPSC-IN2P3, 53, avenue des Martyrs, 38000 Grenoble, France
    \label{LPSC}
    \and
    Kavli Institute for Astrophysics and Space Research, Massachusetts Institute of Technology, Cambridge, MA 02139, USA
    \label{MIT}
}

\abstract{%
    One of the key elements needed to perform the cosmological exploitation of a cluster survey is the relation between the survey observable and the cluster masses.
    Among these observables, the integrated Compton parameter $Y$ is a measurable quantity in Sunyaev-Zeldovich (SZ) surveys, which tightly correlates with cluster mass.
    The calibration of the relation between the Compton parameter $Y_{500}$ and the mass $M_{500}$ enclosed within radius $R_{500}$ is one of the scientific goals of the NIKA2 SZ Large Program (LPSZ).
    We present an ongoing study to forecast the constraining power of this program, using mock simulated datasets that mimic the large program sample, selection function, and typical uncertainties on $Y_{500}$ and $M_{500}$.
    We use a Bayesian hierarchical modelling that enables taking into account a large panel of systematic effects.
    Our results show that the LPSZ can yield unbiased estimates of the scaling relation parameters for realistic input parameter values.
    The relative uncertainties on these parameters is $\sim 10\%$ for the intercept and slope of the scaling relation, and $\sim 34\%$ for its intrinsic scatter, foreshadowing precise estimates to be delivered by the LPSZ.
}
\maketitle
%
\section{Introduction}
\label{sec:intro}

The abundance of galaxy clusters in mass and redshift is tightly linked to large scale structure formation physics.
As a consequence, cluster surveys can be used to constrain cosmological parameters, provided cluster masses can be estimated (see \eg\ \cite{allen_cosmological_2011}).
Such studies therefore rely on a prior knowledge of the link between cluster masses and survey observables, often taking the form of mass-observable scaling relations (SR).
In particular, catalogues of galaxy clusters detected through the Sunyaev-Zeldovich (SZ) effect may use the observable integrated Compton parameter $Y$ as a mass proxy for cosmological studies (\eg\ \cite{planck_collaboration_planck_2016}).

Several approaches can be considered to estimate mass-observable scaling relations.
One of them is to follow-up representative samples of galaxy clusters with dedicated observations, enabling individual mass measurements that can be used to study their correlation with survey observables.
The NIKA2 SZ Large Program (hereafter LPSZ, \cite{mayet_cluster_2020, perotto_nika2_2021}) consists in a high-resolution SZ follow-up of $\sim50$ galaxy clusters selected from SZ surveys.
The combination of NIKA2 SZ observations and of X-ray follow-ups gives access to individual hydrostatic mass measurements for each of these clusters, and thus to studies of the $Y-M$ scaling relation.

In this paper, we present a study forecasting the ability of the NIKA2 LPSZ to constrain the $Y-M$ scaling relation.
We use a Monte-Carlo approach to generate mock realistic LPSZ-like samples with a fiducial scaling relation.
The estimation of the scaling relation from these samples then allows us to forecast typical biases and uncertainties in the analysis.

\section{Scaling relation modelling}

Self-similar scenarii of structure growth predict a power law relation between the thermal energy content of a galaxy cluster and its mass.
The former can be probed by the integrated Compton parameter, defined as:
\begin{equation}
    \label{eq:y500}
    Y_{500} = 4\pi \frac{\sigma_\textsc{t}}{m_{\rm e} c^2} \int_0^{R_{500}} r^2 P_{\rm e}(r) \, {\rm d}r,
\end{equation}
where $\sigma_\textsc{t}$ is the Thomson scattering cross-section, $P_{\rm e}$ the electron pressure in the intracluster medium (ICM), and $R_{500}$ is a characteristic radius of the cluster, defined as the radius of a sphere around the cluster enclosing an average density 500 times greater than the critical density of the Universe at the redshift of the cluster. \\
The $Y-M$ scaling relation can then be written as (\eg\ \cite{kravtsov_new_2006}):
\begin{equation}
    \label{eq:sz_scaling}
    E^{-2/3}(z) \left[\frac{Y_{500}}{10^{-4} \,{\rm Mpc^2}}\right] = 10^\alpha \left[\frac{M_{500}}{3 \times 10^{14} \,M_\odot}\right]^{\,\beta},
\end{equation}
where $M_{500}$ is the mass contained within $R_{500}$, and $E(z) \equiv H(z)/H_0$ is the reduced Hubble parameter at the redshift of the cluster. \\
Defining log-scaled observable and mass values as $Y$ and $Z$ respectively, eq. (\ref{eq:sz_scaling}) becomes a linear relation:
\begin{equation}
    Y = \alpha_\yz + \beta_\yz Z
\end{equation}

The complex physical processes occurring inside galaxy clusters make this linear relation a trend rather than a deterministic relation.
The scaling relation can then be expressed as the probability of a cluster having an observable value given its mass:
\begin{equation}
    \label{eq:yz}
    P(\yz) = \mathcal{N}(\alpha_\yz + \beta_\yz Z, \sigma_\yz^2),
\end{equation}
where $\sigma_\yz$ is the Gaussian intrinsic scatter around the relation.

\section{Mock cluster sample generation} \label{sec:sample_gen}

In order to evaluate the NIKA2 LPSZ constraining power on the $Y-M$ scaling relation, we generate realistic mock cluster samples, with the same properties as the actual LPSZ sample in the $(Y_{500}, M_{500})$ plane.
The LPSZ is composed of 45 clusters between redshifts $0.5$ and $0.9$ \cite{mayet_cluster_2020, perotto_nika2_2021}.
The clusters were selected from \textit{Planck} and ACT SZ catalogues \cite{planck_collaboration_planck_2016-1, hasselfield_atacama_2013} according to their integrated Compton parameter $Y_{500}$.
Five bins in $Y_{500}$ and two bins in redshift were defined, dividing the observable-redshift plane in ten boxes (see bottom right panel of figure~\ref{fig:sample_gen}).
Five clusters were then selected in each of these boxes\footnotemark, creating a relatively homogeneous coverage of the mass range of interest.
Assuming the $Y_{500}-M_{500}$ scaling relation from \cite{arnaud_universal_2010}, the mass range covered roughly spans across $M_{500} \in [3, 11] \times 10^{14} M_\odot$.
\footnotetext{
    There were not enough high-mass, high-$z$ clusters in the \textit{Planck} and ACT catalogues to fill the two highest mass boxes in the LPSZ high-redshift bin.
    This is due to the fact that high mass, high redshift clusters are rare objects in the Universe.
    The resulting sample therefore only includes 45 clusters.
}

The generation of a cluster sample is performed through the following steps, summarized in figure~\ref{fig:sample_gen}:

\begin{figure*}[t]
    \centering
    \includegraphics[width=.95\linewidth]{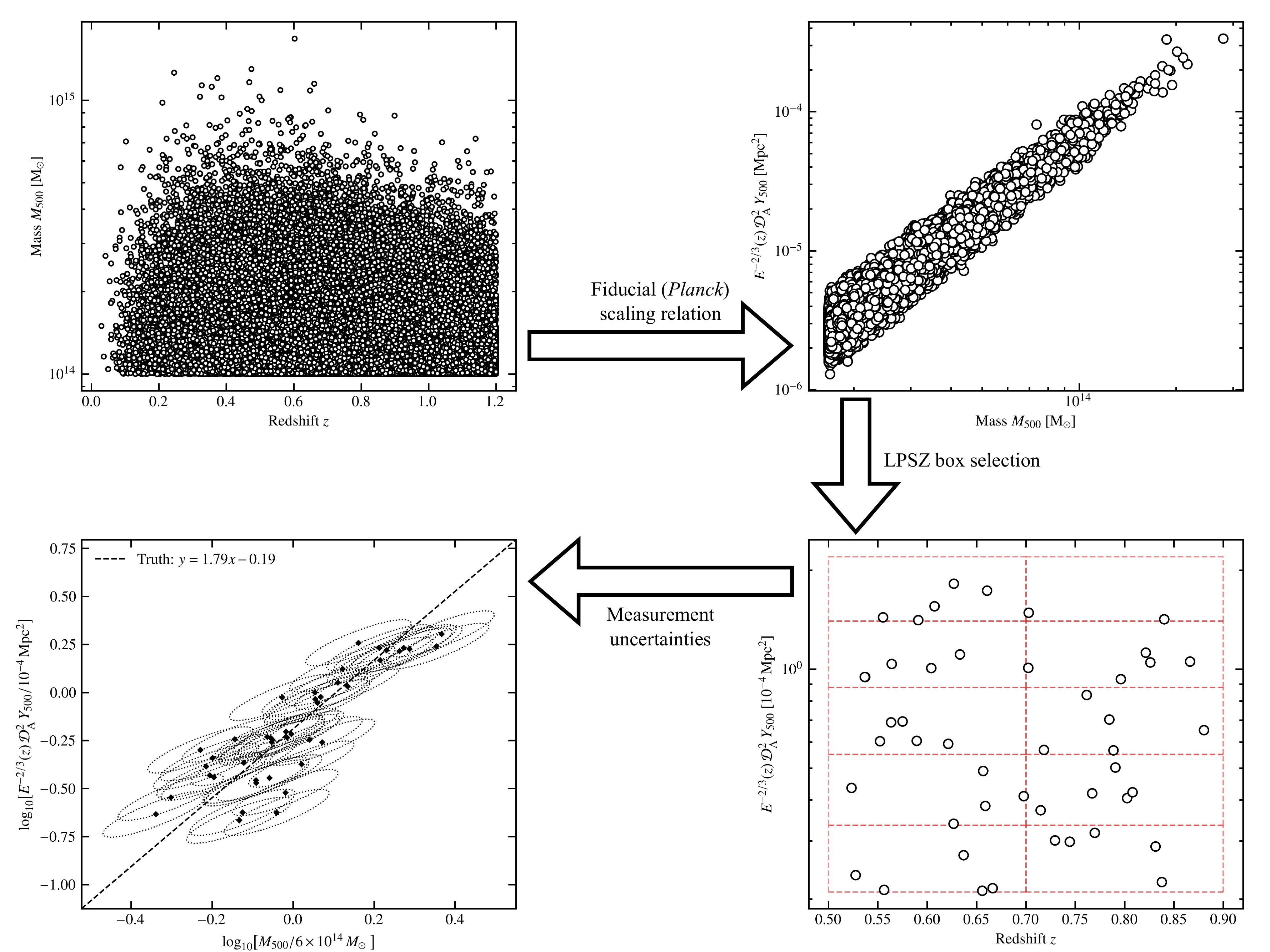}
    \caption{
        Schematic representation of the workflow used to generate a mock cluster sample.
        The produced samples mimic the NIKA2 LPSZ in terms of mass and redshift coverage, as well as data quality.
    }
    \label{fig:sample_gen}
\end{figure*}

\begin{enumerate}[leftmargin=0.5cm]
\item A large number of points is drawn randomly in the $(z, M_{500})$ plane from a Tinker halo mass function \cite{tinker_toward_2008}.
    Given the steepness of the mass function at low masses, we set a lower limit as $M_{500} \geqslant 10^{14} M_\odot$.
    This allows samples to cover a mass range larger than that of the NIKA2 LPSZ.
    The resulting sample is a Universe-like distribution of galaxy clusters in the mass-redshift plane.

\item A fiducial scaling relation is used to compute observable values for each of these clusters.
    We use eq. (\ref{eq:yz}) with parameter values obtained by the \textit{Planck} collaboration \cite{planck_collaboration_planck_2016}:
    \begin{equation}
        \label{eq:true}
        \alpha_\yz^{\rm true} = -0.19 \quad ; \quad
        \beta_\yz^{\rm true} = 1.79   \quad ; \quad
        \sigma_\yz^{\rm true} = 0.075.
    \end{equation}

\item Galaxy clusters are selected according to their SZ observable values and redshift similarly to the NIKA2 LPSZ selection function, \ie\ by randomly picking 5 clusters for each box according to their redshift and $Y_{500}$ value.

\item Correlated uncertainties on the (log scaled) mass and observable are added to each data point.
    We generated realistic mock NIKA2 observations of simulated clusters, with various data quality.
    Processing these maps with the \texttt{PANCO2} software \cite{keruzore_panco2_2021} showed that realistic data quality for NIKA2 LPSZ cluster observations yielded relative uncertainties between $10\%$ and $15\%$ on $Y_{500}$ and $M_{500}$, with a $\sim 85\%$ correlation coefficient.
    These results are consistent with previous NIKA2 studies of galaxy clusters (\eg\ \cite{ruppin_first_2018, keruzore_exploiting_2020}).
\end{enumerate}

These steps are repeated to generate 5000 cluster samples similar to that of the NIKA2 LPSZ, with realistic values of mass and SZ observable, and a known scaling relation between the two.
The estimation of the scaling relation parameters by regression on the mock observed data will then allow us to assess biases and scatter of these estimators, \ie\ the accuracy and precision expected for the LPSZ scaling relation estimation.

It is important to note that the NIKA2 LPSZ sample creation includes one more step, that is not replicated in this study.
Galaxy clusters from the LPSZ sample are selected from cluster catalogues detected by the \textit{Planck} and ACT SZ surveys \cite{planck_collaboration_planck_2016-1, hasselfield_atacama_2013} rather than directly from the true cluster population in the Universe.
By bypassing this step, we ignore the selection function of these two surveys, which is equivalent to assuming that the \textit{Planck} and ACT survey catalogues are a good proxy of the underlying cluster population in the Universe in the portion of the mass-redshift planed covered by the LPSZ.
The validity of this hypothesis and its impact on the scaling relation are to be assessed in a future study.

\section{Scaling relation regression}

\subsection{Regression scheme}

We use the LIRA \texttt{R} software \cite{sereno_bayesian_2016} to estimate scaling relation parameters from our samples.
LIRA provides a ready-to-use regression scheme using a Bayesian hierarchical modelling of the scaling relation, with very flexible parameters and options.
The Bayesian hierarchical approach enables taking into account multiple systematic effects at play in mass-observable scaling relations, such as scatter and bias in the mass estimators, correlated uncertainties on both axes of the relation, or selection effects.
For more detailed information on the use of Bayesian hierarchical modelling for cluster mass-observable scaling relations, we refer the reader to the works of \eg\ \cite{sereno_bayesian_2016, mantz_gibbs_2016}.

LIRA uses a Gibbs sampling Monte Carlo Markov Chains (MCMC) algorithm to sample the posterior probability of the model parameters given the observed data.
Its direct products are therefore Markov chains forming a sampling of the posterior distribution in the parameter space, which can be used to infer the probability distribution for the parameters of interest $(\alpha_\yz, \beta_\yz, \sigma_\yz)$ from eq. (\ref{eq:yz}).
We run LIRA to fit the scaling relation for each of the 5000 realistic mock cluster samples generated in \S\ref{sec:sample_gen}.
For each mock sample, we estimate the bias $\xi$ and the dispersion $\eta$ of each parameter of interest in percent of the true parameter value from the Markov chains:
\begin{equation}
    \xi_\vartheta \equiv 100 \times \frac{{\rm Med} \big[\vartheta_i\big]_i - \vartheta^{\rm true}}{|\vartheta^{\rm true}|}, \qquad
    \eta_\vartheta \equiv 100 \times \frac{\sqrt{{\rm Var}\big[\vartheta_i\big]_i}}{|\vartheta^{\rm true}|}
    \label{eq:scaling:xi_eta}
\end{equation}
where $\vartheta_i$ is the $i$-th sample of the Markov chain for the parameter $\vartheta$, and $\vartheta^{\rm true}$ is the fiducial value used to generate the cluster sample, given in eq. (\ref{eq:true}).
${\rm Med}[\dots]_i$ et ${\rm Var}[\dots]_i$ respectively denote the median and the variance of the Markov chains.
We also define the significance of the bias of each parameter estimator $\zeta$ as:
\begin{equation}
    \label{eq:scaling:zeta}
    \zeta_\vartheta \equiv \frac{{\rm Med} \big[\vartheta_i\big]_i - \vartheta^{\rm true}}{\sqrt{{\rm Var}\big[\vartheta_i\big]_i}}
    = \frac{\xi_\vartheta}{\eta_\vartheta}.
\end{equation}
As a result, we obtain a value of $\xi, \zeta, \eta$ for each of the three parameters of interest for each mock LPSZ sample.

\subsection{Results}

\begin{figure*}[t]
    \centering
    \includegraphics[width=.49\linewidth]{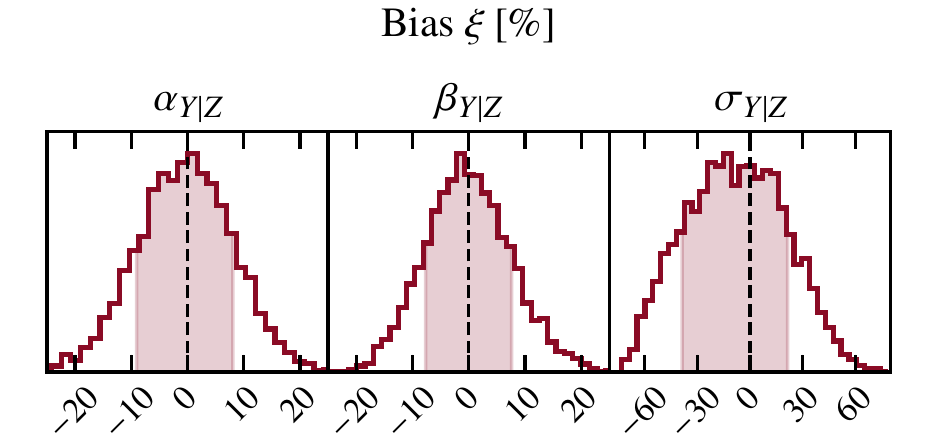}
    \includegraphics[width=.49\linewidth]{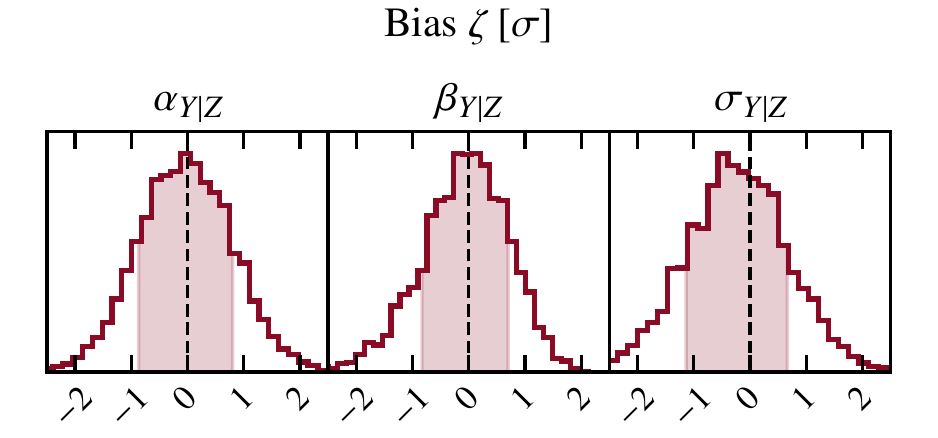} \\[10pt]
    \includegraphics[width=.49\linewidth]{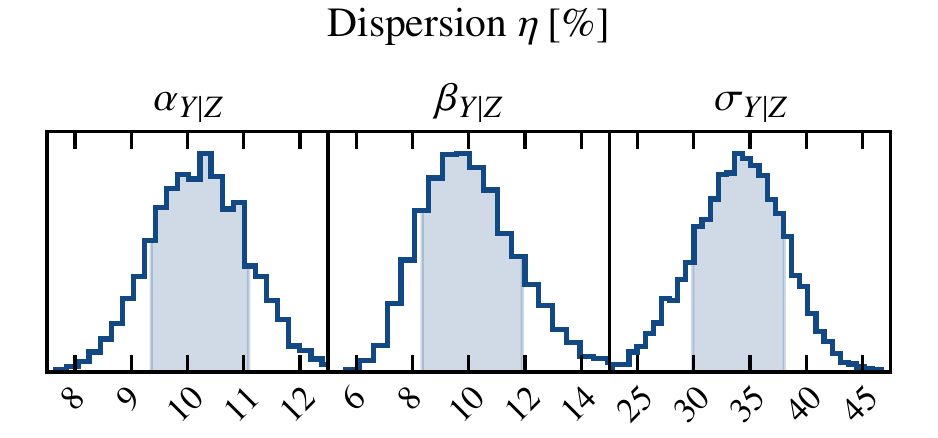}
    \caption{
        Distributions of the biases (\textit{top}) and dispersions (\textit{bottom}) of the scaling relation parameter estimators over the 5000 mock LPSZ-like samples.
        No significant bias is found on the parameters of interest.
        Relative uncertainties are on average $\sim 10\%$ on the slope and intercept of the scaling relation, and $\sim 34\%$ on its intrinsic scatter.
        Shaded regions are $1\sigmaup$.
    }
    \label{fig:results}
\end{figure*}

We present the distributions of the biases and dispersions obtained for our 5000 LPSZ-like samples in figure~\ref{fig:results}.
The top panel shows that the distributions of $\xi$ and $\zeta$ are centred around zero, indicating no bias on average on the parameters of interest of the scaling relation.
This shows that an unbiased estimation of the $Y_{500}-M_{500}$ scaling relation can be recovered from the NIKA2 LPSZ follow-up of galaxy clusters, within the scope of our assumptions.
The bottom panel shows the distribution of the dispersions $\eta$, representing the uncertainties on the recovered scaling relation parameters from the adjustment.
We see that relative uncertainties on the intercept $\alpha_\yz$ and slope $\beta_\yz$ are on average $\sim 10\%$, while the intrinsic scatter $\sigma_\yz$ is less constrained, with $\sim 34\%$ uncertainty on average.
These results are summarized in table~\ref{tab:results}.

\begin{table}[t]
    \setlength{\tabcolsep}{15pt}
    \small
    \caption{
        Summary of the biases and dispersions on each parameter of interest in our scaling relation analysis.
        Values reported are the mean values and $1\sigmaup$ uncertainties.
    }
    \centering
    \begin{tabular}{c c c c}
        \toprule
        Parameter $\vartheta$ & Bias $\xi_\vartheta \, [\%]$ & Bias $\zeta_\vartheta \, [\sigma]$ & Dispersion $\eta_\vartheta \, [\%]$ \\
        \midrule
        $\alpha_{Y|Z}$ & $-0.5 \pm 8.5$ & $0.0 \pm 0.8$ & $10.2 \pm 0.9$ \\
        $\beta_{Y|Z}$  & $0.0 \pm 7.6$ & $-0.1 \pm 0.8$ & $10.1 \pm 1.8$ \\
        $\sigma_{Y|Z}$ & $-8.2 \pm 28.2$ & $-0.2 \pm 0.9$ & $34.0 \pm 4.0$ \\
        \bottomrule
    \end{tabular}
    \label{tab:results}
\end{table}

\section{Summary and conclusions}

We presented a study forecasting the precision and accuracy expected for the estimation of the $Y_{500}-M_{500}$ scaling relation with the NIKA2 SZ Large Program.
We used Monte-Carlo simulations to create mock LPSZ-like cluster samples by following a sample selection procedure similar to the one used to create the real LPSZ cluster sample, and including realistic uncertainties on the measured mass and integrated Compton parameter of each cluster.
We used the LIRA software to adjust a power-law scaling relation between $Y_{500}$ and $M_{500}$.
The comparison of LIRA results with the true fiducial scaling relation used to generate the mock samples did not allow us to identify any significant bias in the analysis.
Moreover, it allowed us to forecast the expected uncertainties on the LPSZ estimates of the scaling relation parameters.
We obtain typical uncertainties of around $10\%$ of the true values for the slope and intercept of the relation, and $34\%$ for its intrinsic scatter.
Such uncertainties are comparable with those obtained through the exploitation of earlier cluster samples using multi-wavelength data (such as \eg\ \cite{sereno_comalit_2015}).
As a combination of high angular resolution SZ and X-ray follow-ups, the NIKA2 SZ Large Program can therefore be expected to deliver quality cluster mass calibrations for SZ surveys.

The two main caveats of this study reside in its hypotheses.
First, when generating mock cluster samples, we neglected selection effects in the \textit{Planck} and ACT surveys.
The inclusion of these effects will be needed to be able to state that the LPSZ can recover the $Y_{500}-M_{500}$ scaling relation of the true cluster population in the Universe.
Second, we have chosen to ignore bias and scatter in the mass estimator used to weigh clusters in the LPSZ.
The combination of SZ and X-rays that will be used allows us to access the hydrostatic mass of clusters, which is well known to be a biased, but low-scatter estimator.
As such, this study focuses on the scaling relation between integrated Compton parameter and the hydrostatic mass of galaxy clusters.
This relation needs to be coupled with external information on the value of the hydrostatic mass bias to be exploitable for cosmological purposes, enabling the propagation of the uncertainty on this bias to cosmological results.
These two caveats will be addressed in a future study.

\section*{Acknowledgements}
\small

This work is supported by the French National Research Agency in the framework of the ``Investissements d’avenir'' program (ANR-15-IDEX-02).
291294).
F.R.  acknowledges financial supports provided by NASA through SAO Award Number SV2-82023 issued by the Chandra X-Ray Observatory Center, which is operated by the Smithsonian Astrophysical Observatory for and on behalf of NASA under contract NAS8-03060.

\bibliography{./mmu_scaling.bib}

\end{document}